\documentclass[preprint2]{aastex}
\usepackage{natbib}

\newcommand{\obj}{WISE 1118+31}
\newcommand{\xiUMa}  {\mbox{$\xi$ UMa}}

\newcommand{\vs}           {{\it vs.}}
\newcommand{\asec}      {\mbox{$^{\prime\prime}$}}
\newcommand{\um}         {\mbox{$\mu$m}}
\newcommand{\amin}      {\mbox{$^\prime$}}
\newcommand{\be}           {\begin{equation}}
\newcommand{\ee}           {\end{equation}}
\newcommand{\bea}          {\begin{eqnarray}}
\newcommand{\eea}          {\end{eqnarray}}

\slugcomment{submitted to the Astronomical Journal}

\shorttitle{Xi Ursae Majoris Brown Dwarf Companion}
\shortauthors{Wright et al.}

\begin{document}

\title{A T8.5 Brown Dwarf Member of the Xi Ursae Majoris System}

\author{Edward L.\ Wright\altaffilmark{1},
M.\ F.\ Skrutskie\altaffilmark{2},
J.\ Davy Kirkpatrick\altaffilmark{3},
Christopher R.\ Gelino\altaffilmark{3},
Roger L.\ Griffith\altaffilmark{3},
Kenneth A.\ Marsh\altaffilmark{4},
Tom Jarrett\altaffilmark{3},
M.\ J.\ Nelson\altaffilmark{2},
H.\ J.\ Borish\altaffilmark{2},
Gregory Mace\altaffilmark{1},
Amanda K.\ Mainzer\altaffilmark{5},
Peter R.\ Eisenhardt\altaffilmark{5},
Ian S.\ McLean\altaffilmark{1},
John J.\ Tobin\altaffilmark{6},
Michael C.\ Cushing\altaffilmark{7}
\altaffiltext{1}{UCLA Astronomy, PO Box 951547, Los Angeles CA 90095-1547}
\altaffiltext{2}{Department of Astronomy, University of Virginia, Charlottesville, VA, 22904}
\altaffiltext{3}{Infrared Processing and Analysis Center,
California Institute of Technology, Pasadena CA 91125}
\altaffiltext{4}{School of Physics \& Astronomy, Cardiff University, Cardiff CF243AA,
UK}
\altaffiltext{5}{Jet Propulsion Laboratory, California Institute of Technology, 4800 Oak Grove Dr., Pasadena, CA, 91109-8001, USA}
\altaffiltext{6}{National Radio Astronomy Observatory, Charlottesville, VA 22903}
\altaffiltext{7}{Department of Physics and Astronomy, MS 111, University of Toledo, 2801 W. Bancroft St., Toledo\ OH\ 43606-3328}
}

\email{wright@astro.ucla.edu}

\begin{abstract}

The Wide-field Infrared Survey Explorer has revealed a T8.5 brown
dwarf (WISE J111838.70+312537.9) that exhibits common proper motion
with a solar-neighborhood (8 pc) quadruple star system - Xi Ursae
Majoris.  The angular separation is 8.5\amin, and the projected physical
separation is $\approx 4000$ AU.
The sub-solar metallicity and low chromospheric activity of 
$\xi$ UMa A argue that the system has an age of at least 2 Gyr. 
The infrared luminosity and color of the brown dwarf suggests the mass of
this companion ranges between 14 and 38 $M_J$ for system ages of 2 and 8 
Gyr respectively.

\end{abstract}


\section{Introduction}

\begin{figure}[tb]
\plotone{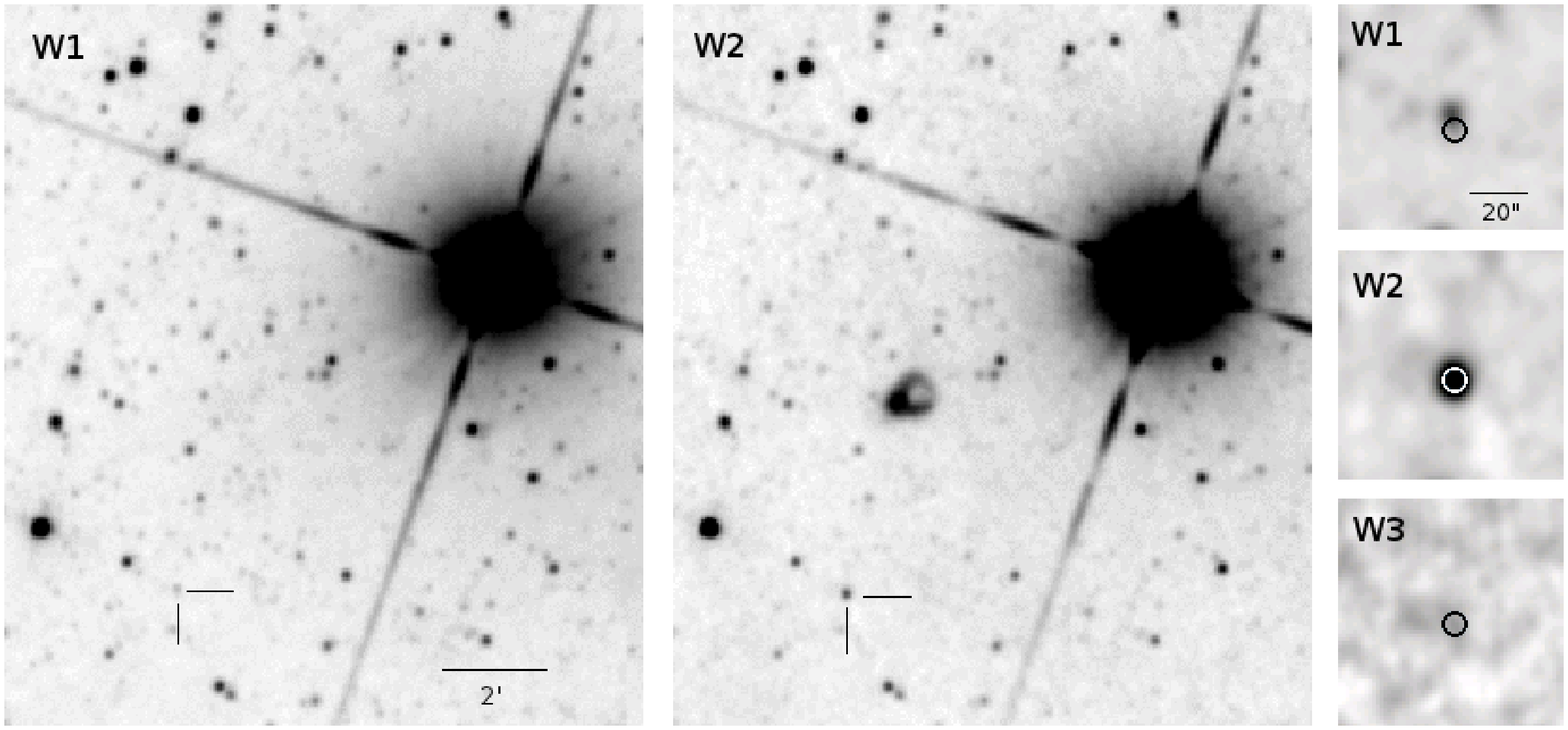}
\caption{WISE All-sky Image Atlas cutouts showing a 10\amin\ FoV
centered halfway between \obj\ (lower left) and \xiUMa\ (upper right).
The left panel shows W1 (3.4 \um), while the right panel shows W2 (4.6 \um).
The  lines point to \obj.  
The faint source near the lines in the left panel is the nearby 2MASS star,
\#1 on Figure \ref{fig:W1118+31-Y}.
The odd shape above 
\obj\ in the W2 image is a ghost image of \xiUMa.
The right-most column shows 1\amin\ postage stamps of WISE images
at 3.4, 4.6 \& 12 \um.
\label{fig:W1118+31-W12}}
\end{figure}

The effective temperature and thus spectrum of a brown dwarf evolves
with time as it cools as a degenerate object \citep{kumar:1962}.  For an
isolated brown dwarf, determination of mass and age are intertwined,
such that a broad locus of mass and age will be consistent with a
single measured effective temperature \citep{burrows/sudarsky/lunine:2003}.  In rare
instances brown dwarfs may reside in close binary systems, resolving
this ambiguity with a direct dynamical mass estimation 
\citep{konopacky/etal:2010, cardoso/etal:2009}.
In the absence of a
dynamically measured mass, spectral modeling can significantly
constrain a brown dwarf's mass if there exists sufficient restriction
on the object's age.  Constraints on age and metallicity are
available if the brown dwarf is a member of a multiple star system.
In this case, the properties of the primary, particularly
chromospheric activity and kinematics, provide an indication of age.

The Wide-field Infrared Survey Explorer mission \citep[WISE,][]{wright/etal:2010} 
has been a productive engine for
the discovery of the coolest brown dwarfs. The WISE  W1(3.4 \um) and
W2 (4.6 \um) filters are optimally
tuned to select the coolest candidates, specifically those with
spectra significantly shaped by methane absorption at low effective
temperature \citep{mainzer/etal:2005}.  To date spectroscopic follow-up of
WISE-selected sources has revealed more than 100 ultracool brown
dwarfs \citep{kirkpatrick/etal:2011} including several exceptionally cool Y-dwarfs
\citep{cushing/etal:2011, kirkpatrick/etal:2012, tinney/etal:2012} demonstrating that WISE colors provide for reliable
photometric selection of ultra-cool brown dwarf candidates.

WISE J111838.70+312537.9, hereafter \obj, easily meets the WISE brown
dwarf color selection criterion \citep{kirkpatrick/etal:2011} with
a W1-W2 color of 2.85 compared to a selection threshold of 2.0.  Few confusing
objects meet these selection restrictions.  
In addition this source lies 8.5\amin\ from one of the nearer stars to the Sun, \xiUMa,
prompting an investigation into a possible system membership. 
This paper reports the spectral characterization of \obj\ and the analysis of a series 
of astrometric observations spanning 26 months aimed at determining whether this 
source exhibits common proper motion with \xiUMa.  These observations demonstrate 
that \obj\ is a newly found member of this already remarkable multiple star system.   
The characteristics of the primary system provide an indication of the metallicity and age of the 
newly discovered ultra-cool brown dwarf constraining the mass of this object.

\section{Observations}

\subsection{WISE}

WISE imaged the region of the sky containing \obj\ on twenty occasions
between 21 May 2010 04:20 UT
and 25 May 2010 09:39 UT. The WISE
All-Sky Catalog reports this source as well detected in W1
and W2, with marginal signal at the position
of the W1/2 source in W3 (12 \um) and only an upper limit in
W4 (22 \um). Of the twenty apparitions, eighteen were sufficiently separated
from a detector edge to permit source extraction.  In all eighteen
cases the source was detected in W2, producing a combined
SNR=34 detection with W2=13.31.  Because exceptionally cool
brown dwarfs are considerably fainter in W1, which was optimized
to produce a substantial flux difference between W1 and W2, \obj\ is detected 
in W1 in only 12 of the 18 opportunities with a combined 
SNR=15 and W1=16.16, yielding a color of W1-W2=2.85.
Figure \ref{fig:W1118+31-W12} shows portions of the WISE image atlas
covering both \obj\ and \xiUMa.

\subsection{Follow-up Imaging}

\begin{table}[tb]
\begin{center}
\caption{Photometric Observations of \obj.\label{tbl:photometry}}
\begin{tabular}{lll}
\tableline\tableline
Filter & Vega Magnitude & Instrument\\
\tableline
Y(MKO) & $19.18\pm0.12$ & FanCam \\
J(MKO) & $17.792\pm0.053$ & WHIRC \\
J &  $ 18.22\pm0.16$ & Bigelow \\
J &  $  18.37\pm 0.08$ & FanCam \\
H(MKO) & $ 18.146\pm0.060$ & WHIRC \\
H & $ 18.13\pm0.23$ & Bigelow \\
Ks(MKO) & $ 18.746\pm0.150$ & WHIRC \\
W1 & $16.160\pm0.071$ & WISE \\
ch1 & $15.603 \pm 0.026$ & IRAC \\
ch2 & $13.368 \pm 0.018$ & IRAC \\
W2 & $13.308\pm0.032$ & WISE \\
W3 & $12.359 \pm 0.314$ & WISE \\
W4 & $ >8.821$ & WISE \\
\tableline
\tablecomments{All magnitudes are Vega magnitudes and use\\ 
the 2MASS JHK$_s$ filters except as noted.
The $Y$-band\\
calibration uses the \citet{hamuy/etal:2006} transformation\\
of $Y-K_s$ \vs\ $J-K_s$.}
\end{tabular}
\end{center}
\end{table}

Multiple epochs of near-infrared imaging provide
the astrometric data for \obj\ needed to confirm common proper motion 
with \xiUMa.  Photometric information from these images in
summarized in Table \ref{tbl:photometry} while the astrometric data
are listed in Table \ref{tbl:astrometry}.   No corrections for non-linearity
were applied to the ground-based photometry because the sky
is brighter \citep{sanchez/etal:2008}
than both \obj\ and the faint 2MASS objects used as calibrators.
The absolute calibration uncertainties of 3\% for Spitzer
\citep{reach/etal:2005} and 2.4, 2.8, 4.5 \& 5.7\%
for W1..4 (WISE Explanatory Supplement \S IV.4.h.v)
relative to Spitzer have not been included in the errors quoted in
Table \ref{tbl:photometry}.

\begin{figure}[tbp]
\plotone{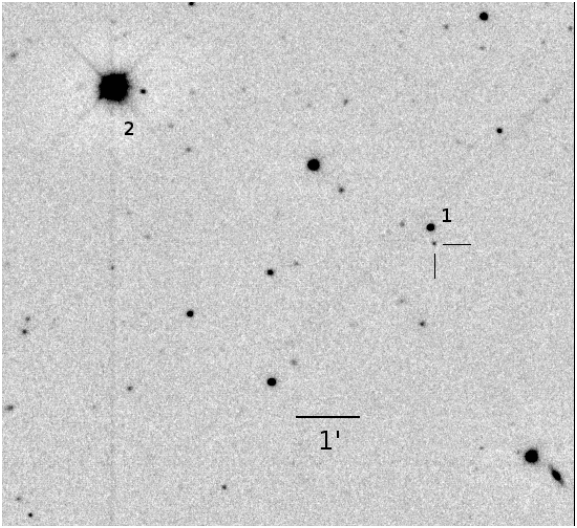}
\caption{Y-band image of the \obj\  field obtained at Fan Mountain
Observatory 15 March 2012 UT.  North is up and East is to the left.
Perpendicular lines mark the position of \obj.  Star 1 nearby is
the J=15.94 star 2MASS J11183876+3125441, while star 2 is 2MASS
J11185086+3126520 with J=10.32.  The faintest detected objects have
$Y \approx 20$.  The faint diffraction spike entering from the upper right
(northwest) is from \xiUMa\  8.5\amin\ away.
\label{fig:W1118+31-Y}}
\end{figure}

\subsubsection{Fan Mountain Observatory/FanCam}

$Y\&J$ photometry and astrometry of \obj\ were obtained at various epochs
between 28 Nov 2010 and 15 Mar 2012 with FanCam, a HAWAII-1 based
near-infrared imager operating at the University of Virginia's Fan
Mountain 31-inch telescope \citep{kanneganti/etal:2009}. 
The source position
was dithered by approximately 10\asec\ between 30 s exposures comprising total
exposure times ranging from 60 to 80 minutes. The FanCam field of view
(FOV) is 8.7\amin\  (0.51\asec/pixel). 
The central $7\amin \times 7\amin$  (Figure \ref{fig:W1118+31-Y})
of the combined, dithered exposures
was fully covered, providing several 2MASS stars for photometric and
astrometric reference. A median sky frame was subtracted from each
individual exposure prior to flat fielding with the median background
level subsequently restored to the image.  
$Y\&J$ aperture photometry was
computed using an aperture with a radius of 3 pixels. The zero points
for the $J$-band were computed using stars in the FOV with measured
2MASS magnitudes since the $J$-band filter in FanCam are based on the
2MASS system. In order to derive the $Y$-band zero point, we first
computed the $Y$-band magnitudes of stars in the FOV using their 2MASS
$J$ and $K_s$ magnitudes and the transformation given by \citet{hamuy/etal:2006}. 
The final uncertainty in the magnitudes include the photon
noise from the sky and source, the read noise, and the uncertainty in
the zero point due to the computed $Y$-band magnitudes of the
calibrators. The resulting magnitudes and uncertainties
are given in Table \ref{tbl:photometry}.

\subsubsection{Mount Bigelow/2MASS}

The former 2MASS camera on the 1.54\,m Kuiper Telescope on Mt. Bigelow,
Arizona, has three 256$\times$256-pixel NICMOS3 arrays
simultaneously observing in 2MASS $J$, $H$, and $K_s$ filters (Milligan et
al. 1996). The plate scale for all three arrays is 1.65\asec/pixel,
resulting in a 7\amin\  field of view. Exposures of 10s duration,
216 in all, of \obj\ were obtained on 20 May 2011 using 6 repeats of
a $3\times3$ box dither pattern,
with four consecutive images taken at each of the nine dither positions.  The data
were reduced using custom IDL routines implementing standard near-infrared
flat fielding, background removal, and co-addition techniques. 
Flat fields in each band were constructed using on-source frames.   
2MASS stars provided photometric reference in all three bands, leading
to the magnitudes reported in Table \ref{tbl:photometry}.

\subsubsection{WIYN/WHIRC}

$JHK_s$ broad-band imaging of WISE 1118+31 was obtained on UT 31 Dec 2011 with the
WIYN High-Resolution Infrared Camera \citep[WHIRC,][]{meixner/etal:2010} 
and the WIYN 3.5-m Observatory.
The data quality is excellent: both seeing ($\sim0.5\asec$ with 0.1\asec\ pixel scale)
and photometric stability conditions were optimal.
For each band, individual frames had exposure times of
120, 120 and 40 seconds, for $J$, $H$ and $K_s$, respectively.
WHIRC uses MKO filters for $JHK_s$.
Using an efficient on-array dither pattern, a total of 7, 9 and 43 frames for $J$, $H$ and $K_s$, respectively,
were obtained, thus providing a total exposure time of
840, 1080 and 1720 seconds for the $J$, $H$ and $K_s$ mosaics respectively.

Individual frames were corrected for pupil-ghosts, dark and median sky flat subtracted,
normalized by dome flat, and distortion corrected using information
from the WHIRC user information guide.
Astrometric and photometric solutions using 2MASS standards were then found.
The fully reduced frames were combined into a deep mosaic, with outlier (bad pixel) rejection
applied using temporal statistics.
A final astrometric and flux calibration was then applied
to the deep mosaic. 
The photometric uncertainty (comparing with the 2MASS PSC) was typically better 
than 5\% for each mosaic produced.
The achieved spatial resolution was $\sim0.5-0.7\asec$ for the final mosaics,
and the astrometric uncertainty was $\sim0.05\asec$.
The target source, \obj, was detected in all three bands. 
Using a 1.1\asec\  radius circular aperture, the background-subtracted integrated 
(Vega) magnitudes found are reported in Table \ref{tbl:photometry}.
Note that WHIRC uses MKO filters, and the expected difference 
$J_\mathrm{2MASS}-J_\mathrm{MKO}$ for a T8.5 brown dwarf
is $\approx0.36$ mag \citep{stephens/leggett:2004}.  If we correct the
$J_\mathrm{2MASS}$ values by this amount we get three values for
$J_\mathrm{MKO}$: 17.79, 17.86 \& 18.01.  The mid-range of these
values in $J_\mathrm{MKO} = 17.9$ which we adopt, giving $J_\mathrm{MKO}-W2
=4.6$.

\subsubsection{Spitzer}

The Infrared Array Camera (IRAC) \citep{fazio/etal:2004} onboard the {\it  
Spitzer} Space Telescope employs 256$\times$256-pixel detector arrays  
to image a field of view of 5$\farcm$2$\times$5$\farcm$2 (1$\farcs$2  
pixel$^{-1}$).  IRAC was used during the warm {\it Spitzer} mission to  
obtain deeper photometry in its 3.6 and 4.5 $\mu$m channels  
(hereafter, ch1 and ch2, respectively) than WISE was able to take in  
its W1 and W2 bands.
These observations  
were made as part of Cycle 7 and Cycle 8 programs 70062 and 80109  
(Kirkpatrick, PI).  Our standard data acquisition  
and reduction methodology for IRAC observations is outlined in  
\citet{kirkpatrick/etal:2011}.

\begin{figure}
\epsscale{1.0}
\plotone{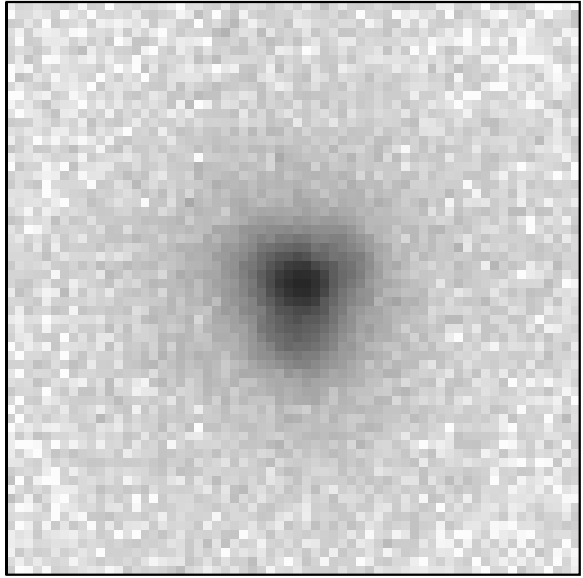}
\caption{$H$-band LGS-AO image of WISE 1118+3125 taken with the NIRC2 camera on 
Keck II.  The image is $\approx$0$\farcs$6 on a side with North up and East to the 
left.  There is no evidence for an equal brightness companion beyond a separation
of 50 mas.\label{fig:w1118_mosaic}}
\end{figure} 

\subsubsection{Keck Adaptive Optics Imaging}\label{sec:AO}

High resolution imaging observations of \obj\ were obtained using the 
Keck II LGS-AO system \citep{wizinowich/etal:2006,vandam/etal:2006} with NIRC2 on 
14 April 2012 (UT).  WISE 1118+3125 is not bright enough to serve as the tip-tilt
reference star for the LGS-AO system, so we used the nearby 2MASS star (\# 1
on Figure \ref{fig:W1118+31-Y}),
which has $R$=16.7 and is located about 7\asec\ from the target.
The seeing was generally very good throughout the night 
(0$\farcs$3--0$\farcs$5, however high clouds were present during the 
observations of  \obj.  We used the MKO $H$ filter and narrow
plate scale (0.009942$\asec$/pixel for a single-frame field-of-view of 
$10\asec\times10\asec$) for the observations.  The data were obtained 
by using a 3-point 
dither pattern that avoided the noisy, lower left quadrant of the array.  Each 
image had an integration time of 120 s and the dither pattern was repeated five
times to give a total exposure time of 1800 s.

The images were reduced in a standard fashion using custom IDL scripts.  These 
steps included dark frame subtraction, flat fielding (using a dome flat), and 
sky subtraction from a sky frame created from the dithered science frames.  The
individual frames were then shifted to move \obj\  to the center of 
the array and the stack was median averaged to create the final mosaic seen 
in Figure~\ref{fig:w1118_mosaic}.  The FWHM in the final mosaic is 54 mas and
shows no irregularities or evidence for a close companion.

\subsection{Spectroscopy}\label{sec:spectroscopy}

\subsubsection{LBT-LUCIFER}

We obtained an $H$- and $K$-band spectrum of \obj\ on 2012 Dec 12 (UT)
using the Large Binocular Telescope (LBT) Near-Infrared Spectroscopic
Utility with Camera and Integral Field Unit for Extragalactic Research
\citep[LUCIFER,][]{LUCIFER:2008}.   
A series of twelve 300 s exposures was obtained at different positions
along the 4\amin\ slit to facilitate sky subtraction.  The A0 V
star HD 97034 was also observed for telluric correction and flux
calibration purposes.  A series of halogen lamp exposures were also
obtained for flat fielding purposes.

The data were reduced using custom Interactive Data Language (IDL)
software based on the Spextool data reduction package
\citep{cushing/vacca/rayner:2004}.  Pairs of images taken at two different
positions along the slit were first subtracted and flat-fielded.  The
spectra were then extracted and wavelength calibrated using sky emission
lines of OH
and CH$_4$.  The twelve spectra are combined 
and then corrected for telluric absorption and flux calibrated using the
technique described by \citet{vacca/cushing/rayner:2003}.  
The final spectrum is shown in Figure  \ref{fig:W1118-spectrum}.

\subsubsection{Hale - TripleSpec}

\begin{figure}[tbp]
\plotone{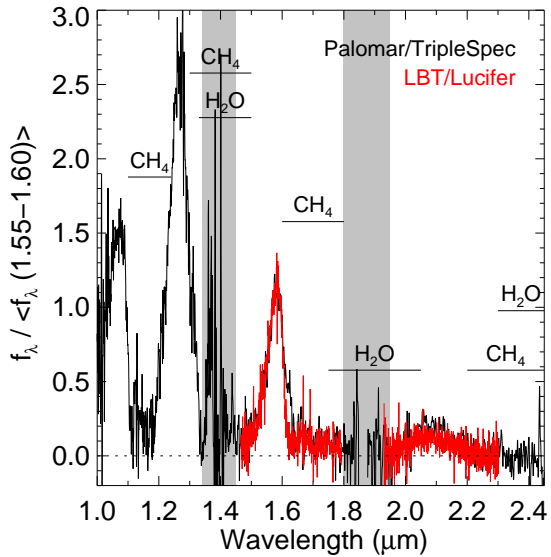}
\caption{\label{fig:W1118-spectrum}Spectrum of \obj\ obtained with TripleSpec
  (\textit{black}) and LUCIFER (\textit{red}).  
  Regions of strong telluric absorption are shown in grey.
  Prominent absorption
  bands of CH$_4$ and H$_2$O are indicated.  The agreement
  between the two spectra is excellent.}
\end{figure}

A 1$-$2.5 $\mu$m spectrum of \obj\ was obtained with the Triple
Spectrograph \citep[Triplespec,][]{herter/etal:2008} at the 5.08\,m Hale
Telescope at Palomar Observatory.  The 1$-$2.5 $\mu$m range is covered
over four cross-dispersed orders which are imaged simultaneously on the
1024 $\times$ 2048 HAWAII-2 array.  The 1\asec-wide slit provides a
resolving power of $R \approx 2700$.  A series of eight, 300 s
exposures were obtained at two different positions along the 30\asec-long
slit to facilitate sky subtraction.  The A0 V star HD 99966 was observed
for telluric correction and flux calibration purposes and dome flats
were obtained at the start of the night.

The data were reduced using a modified version of the Spextool
\citep{cushing/vacca/rayner:2004} package; a detailed description of the
reduction steps can be found in \cite{kirkpatrick/etal:2011}.  Briefly, a
two-dimensional wavelength solution is derived using sky emission
features of OH and CH$_4$.  Spectra are then extracted from
pair-subtracted, flat-fielded images. The resulting spectra are combined
and corrected for telluric absorption and flux calibrated on an
order-by-order basis.  Finally, the spectra from each order are stitched
together to form a complete 1$-$2.5 $\mu$m spectrum.  The spectrum was
then flux calibrated as described in \cite{rayner/cushing/vacca:2009} using
the photometry in Table 1.  The final spectrum is shown in 
Figure \ref{fig:W1118-spectrum}.

\section{Discussion}

\subsection{Properties of  the Central Star System}

$\xi$ Ursae Majoris\footnote{also Alula Australis, HR 4374/5, Gl 423, 
HD 98230/1}  is a complex stellar system with 
at least 4, and possibly 5 components 
\citep{mason/etal:1995} known prior to the discovery of \obj.
Visible to the unaided eye a short distance from the Big Dipper, 
this telescopic double star was among the first to be
recognized as a gravitationally bound binary system \citep{herschel:1804}.
Given the 60 year period of the visual pair it was not until 23 years
later that \citet{struve:1827} calculated a formal orbit.  Subsequently
both components of the visual pair with $a = 2.536\asec$ 
\citep{mason/etal:1995}  were found to be
spectroscopic binaries.
\citet{heintz:1967} finds periods of 669.1 days for the Aa system and 3.9805 days
for the Bb system.
\citet{griffin:1998} gives velocity amplitudes, $K_{Aa} = 4.85\pm0.14$ km/sec
and $K_{Bb} = 4.33\pm0.09$ km/sec in the 60 year orbit
which lead to a dynamical parallax estimate
of $0.126\asec \pm 0.0023\asec$.

However \citet{soderhjelm:1999} re-analyzed the Hipparcos data combined
with other data for visual binaries and gives $0.1197\asec \pm 0.0008\asec$
for the parallax of \xiUMa, and total mass for the AaBb system of 2.62 $M_\odot$.

The weighted mean of the dynamical parallax and the Hipparcos parallax
is $0.1206\asec \pm 0.00074\asec$ but with
$\chi^2 = 8.6$ for 1 degree of freedom in the fit for the mean, 
so we inflate the errors by a factor of $\sqrt{8.6}$ and adopt 
$0.1206\asec \pm 0.0022\asec$ for the parallax.

Since the possible 5$^{th}$ component, seen by \citet{mason/etal:1995}  
separated by 56 milli-arcseconds from the B component of the visual binary, 
was only detected at one epoch we will not consider it to be a member  of the system.

\citet{bakos/sahu/nemeth:2002} give a proper motion of $0.75\asec$/yr
in position angle $217.49^\circ$, based on a 29 year interval.  These values
resolve into $-0.456\asec$/yr in RA and $-0.595\asec$/yr in Dec which we adopt.

\subsubsection{Spectral Types and Metallicities of the Central Stars}

\citet{cayrel/etal:1994} used high resolution spectroscopy
to find effective temperatures  and gravities of
$5950\pm30$~K,  $\log g = 4.3 \pm 0.2$ for \xiUMa\ A;
and  $5650 \pm 50$~K, $\log g = 4.5 \pm 0.2$ for B.
These temperatures and gravities are consistent with
spectral types of F8.5V and G2V assigned to  \xiUMa\ A
and B by \citet{keenan/mcneil:1989}.
\citet{cayrel/etal:1994}  found that both stars had slightly sub-solar
iron abundances, [Fe/H] $= -0.32 \pm 0.05$~dex.

\subsubsection{Chromospheric Activity and Age of the Central Stars}

\citet{cayrel/etal:1994} see chromospheric calcium emission lines
for \xiUMa\ B but not for \xiUMa\ A.  They propose that the short 
4 day period orbit of the Bb system is driving the chromospheric
activity in B.  From the lack of emission in A, \citet{cayrel/etal:1994} 
conclude that \xiUMa\ is older than stars in the cluster NGC 752, 
which has an age of 2 Gyr.
\citet{ball/etal:2005} report X-ray observations of $\xiUMa$
which show that all of the observed X-ray emission is coming from the
4 day period binary Bb.  X-ray flux from component A is more than
2 orders of magnitude fainter,
implying $\log L_X < 27.5$.  Since component A is close to solar luminosity, 
the ratio of X-ray to bolometric flux is $R_X = \log(L_X/L_{bol}) < -6$ which
gives an age greater than 4 Gyr
using Equation(A3) in \citet{mamajek/hillenbrand:2008}. 
The low space velocity $(U, V, W) = (-2.36 \pm 0.33, -28.45 \pm 1.22,
-20.26 \pm 0.25)$ km/sec calculated by 
\citet{karatas/etal:2004} implies that \xiUMa\ is a member of the
thin disk, so ages much greater than 8 Gyr are unlikely.
Finally
\citet{cayrel/etal:1994} found that the luminosities
and effective temperatures were consistent within the errors
with the masses derived by \citet{heintz:1967} for
a 5 Gyr isochrone calculated with sub-solar abundance.

\subsection{Astrometry of the Companion}\label{sec:astrometry}

Astrometric information was extracted from the observed images at the various
epochs using the standard maximum likelihood technique in which a point spread
function (PSF) was fit to each source profile.  The positional uncertainties
were estimated using an error model which includes the effects of instrumental
and sky background noise and PSF uncertainty.  The PSF and its associated
uncertainty map were estimated for each image individually using a set of
bright stars in the field.  In order to minimize systematic effects, our
astrometry was based on relative positions, using as a reference the nearby
star 2MASS J11183876+3125441 at a separation of approximately 6.5\asec\ 
(star 1 in Figure \ref{fig:W1118+31-Y}). This is
sufficiently close that systematic errors due to such effects as focal-plane 
distortion, plate scale, and rotation errors cancel out in the relative 
position to an accuracy much greater than the random estimation errors.  We do,
however, assume that the proper motion and parallax of the reference star are 
negligible compared to those of  \obj. In support of this assumption, 
we find no significant difference in relative motion of \obj\  when the 
differential astrometry is repeated using more distant 2MASS reference stars, 
at separations of 69\asec\  and 105\asec.  The astrometric estimation procedure is 
discussed in more detail by \citet{marsh/etal:2012}.

Table \ref{tbl:astrometry} gives the measured position offsets which were input
to a parallax and proper motion code that handles both Earth-based and
Spitzer observations.  Table \ref{tbl:astrometry-fits} gives the output from the code for
three different sets of free parameters.  In one case the proper motion 
and parallax were forced to zero.  In the second case they were forced to
match $\xi$ UMa.  In the final case the proper motion and parallax
were left as free parameters.

\begin{table*}[tb]
\begin{center}
\caption{Astrometric Observations of \obj.\label{tbl:astrometry}}
\begin{tabular}{lllll}
\tableline\tableline
Date & \multicolumn{1}{c}{$\Delta\alpha\cos\delta\;[{}\asec]$} & 
\multicolumn{1}{c}{$\Delta\delta \;[{}\asec]$} & Observatory & Band \\
\tableline
2010.39 & $  -0.306\pm   0.162$ & $  -0.444\pm   0.182$ & WISE     &  W   \\
2010.91 & $  -0.165\pm   0.211$ & $  -0.603\pm   0.191$ & FanMt    &  Y   \\
2010.91 & $  -0.208\pm   0.162$ & $  -0.477\pm   0.184$ & WISE     &  W   \\
2011.05 & $  -0.232\pm   0.067$ & $  -0.794\pm   0.067$ & Spitzer  &  ch2 \\
2011.12 & $  -0.414\pm   0.234$ & $  -0.600\pm   0.194$ & FanMt    &  J   \\
2011.24 & $  -0.393\pm   0.113$ & $  -0.765\pm   0.100$ & FanMt    &  J   \\
2011.38 & $  -0.651\pm   0.086$ & $  -0.829\pm   0.084$ & MtBglw   &  J   \\
2011.38 & $  -0.554\pm   0.194$ & $  -0.723\pm   0.190$ & MtBglw   &  H   \\
2011.85 & $  -0.639\pm   0.190$ & $  -1.280\pm   0.170$ & FanMt    &  J   \\
2012.00 & $  -0.650\pm   0.005$ & $  -1.253\pm   0.004$ & WIYN     &  J   \\
2012.00 & $  -0.676\pm   0.010$ & $  -1.235\pm   0.012$ & WIYN     &  H   \\
2012.00 & $  -0.766\pm   0.037$ & $  -1.239\pm   0.038$ & WIYN     &  K   \\
2012.20 & $  -0.850\pm   0.165$ & $  -1.339\pm   0.153$ & FanMt    &  Y   \\
2012.15 & $  -0.695\pm   0.117$ & $  -1.631\pm   0.126$ & Spitzer  &  ch2 \\
2012.52 & $  -1.081\pm   0.136$ & $  -1.611\pm   0.139$ & Spitzer  &  ch2 \\
\tableline
\tablecomments{Offsets relative to 11$^h$ 18$^m$ 38.69$^s$  +31$^\circ$ 25\amin\  37.7\asec\ (J2000).\\
Reference star: 11$^h$ 18$^m$ 38.77$^s$  +31$^\circ$ 25\amin\  44.2\asec\ (2MASS).}
\end{tabular}
\end{center}
\end{table*}

\begin{table*}[tb]
\begin{center}
\caption{Astrometric Fits for \obj.\label{tbl:astrometry-fits}}
\begin{tabular}{llllll}
\tableline\tableline
Type & $\chi^2$ & \#df & $\cos\delta d\alpha/dt$ [{}\asec/yr] &
$d\delta/dt$ [{}\asec/yr] & $\varpi$ [{}\asec] \\
\tableline
Fixed                        & 270.836 & 28 & $\phantom{-}0$ & $\phantom{-}0$ & 0 \\
Forced to $\xi$ UMa & 29.856 & 28 & $-0.456$  & $-0.595$  &  $0.1206$ \\
Free                          & 28.579 & 25 & $-0.419\pm0.048$   & $-0.563\pm0.045$ & $0.124\pm0.033$ \\
\tableline
\end{tabular}

\end{center}
\end{table*}

\begin{figure}[tb]
\plotone{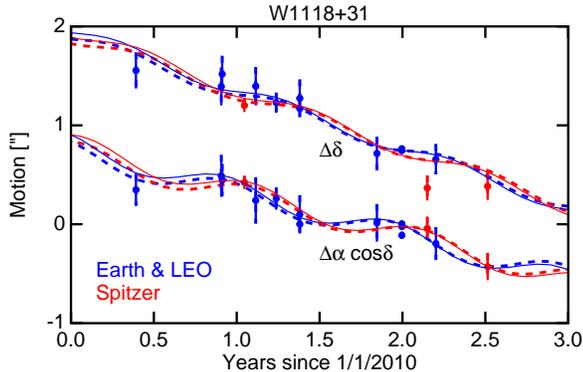}
\caption{Astrometric data and fits for \obj.  
The blue curves and point are for ground-based or Low Earth Orbit observatories,
while the red curves and points are for Spitzer.
The bold dashed lines show the fit with
proper motion and parallax as free parameters, while the lighter solid curves
show the fit forced to match $\xi$ UMa.  These fits are very similar.
Note that $\Delta\delta$ has been displaced by a constant for clarity.
\label{fig:W1118+31_track}}
\end{figure}

The astrometric fits for \obj\ are a very good match to the motion of \xiUMa.
The $\Delta \chi^2 = 241.0$ between the best fit and a fit with a fixed position shows
that the motion of \obj\ has been detected with a SNR of 15.5, while the 
$\Delta \chi^2 = 1.28$ for 3 extra degrees of freedom between the best fit 
and a fit forced to match the proper motion
and parallax of $\xi$ UMa is perfectly consistent with \obj\ being a bound member of
the $\xi$ UMa system.  Figure \ref{fig:W1118+31_track} shows two fits: one forced
to match \xiUMa\  and one with the proper motion and parallax as free parameters.

\subsection{The Nature of the Companion}

\subsubsection{Spectral Classification}

\begin{figure}[tbp]
\plotone{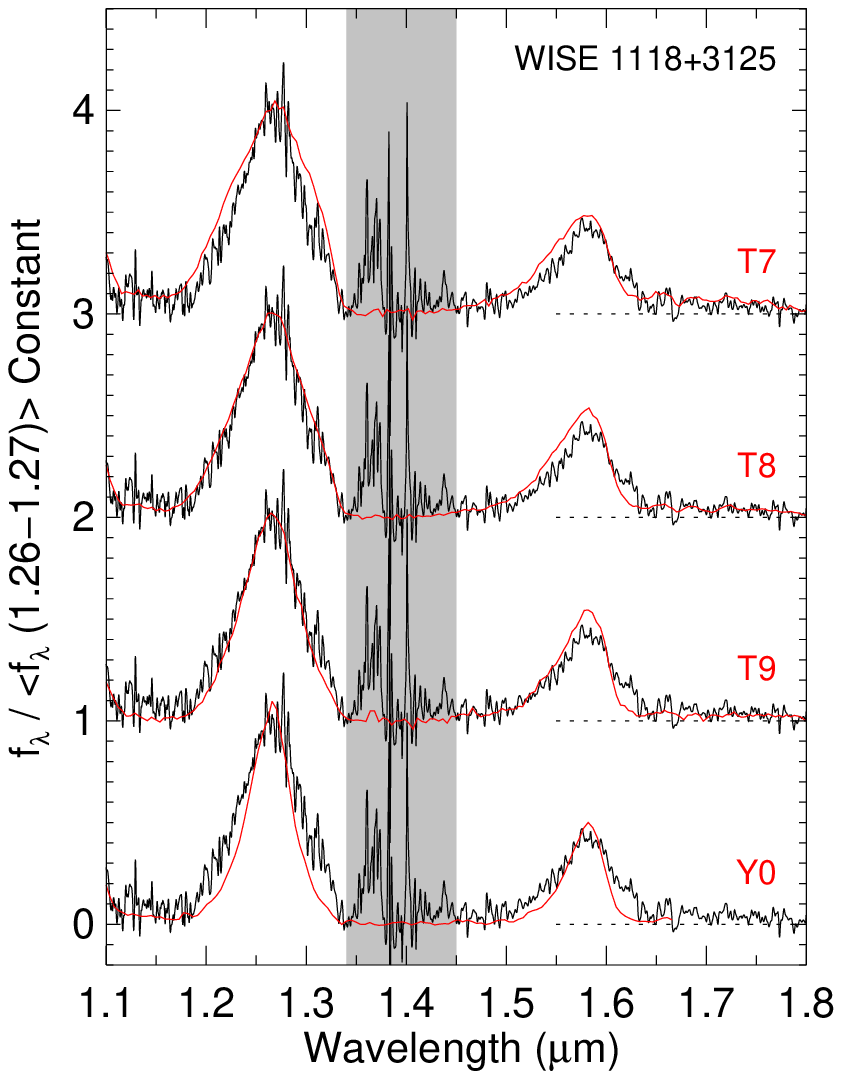}
\caption{Sequence of spectral standards from T7 to Y0 (red) along with 
the spectrum \obj\ (black).  The spectral standards are 
2MASS 0727+1710 \citep[T7,][]{burgasser/etal:2006,burgasser/etal:2002} and 
2MASS 0415-0935 \citep[T8,][]{burgasser/etal:2006,burgasser/etal:2002}, 
UGPS 0722-05  \citep[T9,][]{cushing/etal:2011,lucas/etal:2010}, and 
WISE 1738+2732 \citep[Y0,][]{cushing/etal:2011}.  
The spectrum of WISE 1118+3125 
has been smoothed to a resolving power of $\lambda/d\lambda = 1000$ for display 
purposes.  Spectra are normalized to unity over the 1.26-1.27 \um\  wavelength 
range and offset for clarity (dotted lines).  Regions of high telluric absorption 
are shown in grey.  \obj\ is classified as T8.5$\pm0.5$.
The mismatch between the spectrum of \obj\ and the templates at 
$\sim$1.63 $\mu$m is due to poor subtraction of the OH sky lines.
\label{fig:SpType}}
\end{figure}

As shown in Figure \ref{fig:W1118-spectrum}, 
the spectrum of \obj\ exhibits deep absorption
bands of CH$_4$ and H$_2$O indicative of T dwarfs.  We derive a more
precise spectral type using the spectral classification schemes of
\citet{burgasser/etal:2006} and the extension to this system by
\citet{cushing/etal:2011}
whereby UGPS J072227.51-054031.2 \citep[hereafter UGPS
0722$-$05,][]{lucas/etal:2010} is defined as the T9 spectral
standard and WISE 1738+2732 is defined as the Y0 spectral standard.
As shown in Figure \ref{fig:SpType},
\obj\ has a spectral type of T8.5 based on the width of the J-band peak
at 1.27 \um.

\subsubsection{Absolute magnitude, Luminosity and Mass}

For our adopted parallax
the absolute magnitude is $M_{W2} = W2+5\log(10\varpi) = 13.715 \pm 0.051$.
This agrees very well with
the $M_{W2}$ \vs\ type relation in \citet{kirkpatrick/etal:2011}  which gives
$M_{W2} = 13.71 \pm 0.21$ for a spectral type of T$8.5\pm0.5$.
The $M_{W2}$ \vs\ type relation in \citet{kirkpatrick/etal:2012} with
W1828+2650 excluded gives
$M_{W2} = 13.64 \pm 0.31$ which also agrees with the observed W2 flux.
This $M_{W2} = 13.715$ implies $\nu L_\nu = 1.14 \times 10^{-6}\;L_\odot$
at 4.6 $\mu$m.

To convert $\nu L_\nu$ to $L$ we need the factor 
$\nu F_\nu(W2)/F_\mathrm{bol}$, which is equivalent to the bolometric
correction, since $\nu F_\nu = (\lambda_{iso} F^\circ_\lambda/f_c) 10^{-0.4 m}$
\citep{wright/etal:2010}
which is $1.118 \times 10^{-7}$ erg/cm$^2$/sec for W2=0 with a $\nu^{-1}$
spectrum,
and $F_\mathrm{bol} = 2.48 \times 10^{-5} 10^{-0.4m_\mathrm{bol}}  
\;\mbox{erg/cm$^2$/sec}$,
so
\be
2.5 \log\left(\frac{\nu F_\nu(W2)}{F_\mathrm{bol}}\right) =  
-5.87 +m_\mathrm{bol} - m(W2).
\ee
Thus
\bea
BC(W2) & = & m_\mathrm{bol} - m(W2)\nonumber \\
& = & 2.5 \log\left(\frac{\nu F_\nu(W2)}{F_\mathrm{bol}}\right) +5.87.
\eea

Note that $\nu F_\nu/F_\mathrm{bol}$ is non-negative and also
normalized since $\int (\nu F_\nu/F_\mathrm{bol}) d\ln\nu = 1$ by
definition.  Since the normalization integral is dominated by the
brightest peaks, errors in the faint parts of the spectral energy
distribution (SED) have only small effects on the derived value of
$\nu F_\nu(W2)/F_\mathrm{bol}$ or equivalently the bolometric
correction.
The brightest peaks in $\nu F_\nu/F_\mathrm{bol}$ for late T dwarfs
are in the $J$ band and the $W2$ band, so if we get the $J-W2$ color
right then our value of $\nu F_\nu(W2)/F_\mathrm{bol}$ will be fairly
accurate.

We will use a model spectrum to evaluate $\nu F_\nu(W2)/F_\mathrm{bol}$
in order to fill in the gaps between the photometric passbands,
but we are only using the shape of the spectrum to calculate the
bolometric correction.  
Thus, even if the effective temperature $T_\mathrm{eff}$ is a parameter
of the model, that does not imply that the model parameter is the actual $T_\mathrm{eff}$ of
the star.  

Given a model spectrum $F_\nu$, the ratio $\nu F_\nu(W2)/F_\mathrm{bol}$ is given by
\be
\frac{\nu F_\nu}{F_\mathrm{bol}}
= \frac{\int (\nu F_\nu/h\nu) R(\nu) d\ln\nu}
{\int ([1\;\mathrm{erg/cm}^2]/h\nu) R(\nu)d\ln\nu \; \int F_\nu d\nu}
\ee
where $R(\nu)$ is the response per photon \citep{wright/etal:2010}.
The model we use is a linear combination of sulfide cloud models
from \citet{morley/etal:2012} with $T_\mathrm{eff} = 600$~K, $\log(g) = 5.0$.
We form the sum of 0.2 times the flux from the cloudless model and 0.8 times
the flux from the $f_{sed} = 4$ model.  Physically this is a brown dwarf that
is 20\% covered by clear zones and 80\% covered by cloudy bands.
This linear combination gives $\nu F_\nu/F_\mathrm{bol} = 1.46$,
$J_\mathrm{MKO}-W2 = 4.60$, $(Y-J)_\mathrm{MKO} = 1.23$,
$(J-H)_\mathrm{MKO} = -0.18$, $(J-K_s)_\mathrm{MKO} = -0.06$
and $W1-W2 = 3.17$.  The model matches
the observed $J-W2$ color by design but is too bright in $K_s$
and too faint in $W1$. 
If we make an {\it ad hoc} correction to the SED by making the wavelength range
from 3.0 to 3.8 $\mu$m covering the $W1$ band 0.4 mag brighter that would 
match the $W1-W2$ color and change
the normalization integral by 2\%. 
If we make the wavelength range from
2.0 to 2.3 $\mu$m covering the $K_s$ band 0.8 mag fainter, then the $J-K_s$ color matches the data and the normalization integral changes by -1.3\%.
Thus we adopt $\nu F_\nu(W2)/F_\mathrm{bol} = 1.46$ with a 10\% uncertainty to allow for
such errors in the model, giving
$L = (1.14/1.46)  \times 10^{-6} = 10^{-6.107 \pm 0.043} \;L_\odot$ and
$BC(W2) = 6.28$ so $M_{bol} = 20.0 \pm 0.11$ for \obj.
Interpolating the $L$ \vs\ $M$ and $t$
figures in \citet{saumon/marley:2008} gives this luminosity
for masses of 14 to 31 M$_J$ with ages of 2 and 8 Gyr using the
cloudy models, and 17 to 38 M$_J$ for the cloudless models.

Even though the model parameters do not necessarily apply to the actual star,
we find that  they do give a mass and luminosity consistent with those derived above.
The $T_\mathrm{eff} = 600$~K, 
$\log(g) = 5.0$ models of \citet{hubeny/burrows:2007}
have radii of $0.91\;R_J$, and a mass $M = gR^2/G = 32\;M_J$.
The luminosity calculated directly from $T_\mathrm{eff}$ and $R$ is
$L = 4\pi R^2 T_\mathrm{eff}^4 = 10^{-6.01}\;L_\odot$ 
so this model is reasonably self-consistent.
Calculating the temperature from $R$ and $L$ gives
$T_\mathrm{eff} = (0.91\;R_J/R)^{1/2}(567 \pm 14)$~K.

\subsection{Another Binary?}

If something like the \citet{morley/etal:2012} sulfide clouds is not causing the red
$J-W2$ color of \obj, then
there is a tension between the absolute magnitude $M_{W2}$, which
requires a fairly high $T_\mathrm{eff}$, and the colors and spectral type 
which suggest a lower temperature.  This could be ameliorated if
\obj\ were an equal mass binary which would increase the total radiating
area.  In this case the absolute magnitude
of one component of the binary would be $M_{W2} = 14.47$ and
the colors would remain the same.  The best fitting 
\citet{hubeny/burrows:2007}  model is then a non-chemical equilibrium
model with $T_\mathrm{eff} = 500$~K and $\log(g) = 5.0$.
This model has $M_{W2} = 14.27$, W1-W2 = 4.43, J(MKO)-W2 = 4.57,
and $R = 0.888\,R_J$.
$\nu F_\nu/F_\mathrm{bol} = 1.546$ in the W2 band for this model, so the
luminosity of a single component is $L = 0.5 \times 1.14 \times 10^{-6}/1.546
= 3.7 \times 10^{-7}\;L_\odot$.  The mass range for ages from 2 to 8 Gyr
in the \citet{saumon/marley:2008} cloudless models is 12 to 29 M$_J$,
while the cloudy models give 10 to 25 M$_J$.
Thus the range of total masses is 21 to 59 M$_J$ under the 
equal mass binary hypothesis, which is not much different than the range
under the single object hypothesis.
However, the absolute magnitude of the
components would no longer agree with the $M_{W2}$ \vs\ type relation 
\citep{kirkpatrick/etal:2011,kirkpatrick/etal:2012}.

As stated in \S\ref{sec:AO}, we saw no evidence for binarity in the 14 Apr 2010 
NIRC2 images of \obj.  To determine the upper limit on the separation of an 
equal brightness companion, we added a shifted copy of the mosaic to the 
original mosaic.  At separations $\le$ 4 pixels, we are unable to convincingly 
establish the presence of the companion because of the difficulty in differentiating 
between an irregular PSF caused by an off-axis tip-tilt reference star and
an elongated PSF caused by a companion when no other stars are present on the
array.  At separations of $\ge$5 pixels ($\ge$50 mas), the presence of a 
companion is obvious.  Therefore, we set the
upper limit on the separation to 50 mas, slightly less than the FWHM of the 
image.  If this was not due to a temporary conjunction, so the
actual separation were less than 0.4 AU, then the orbital period could be
less than 2 years.

\subsection{The View from the Neighborhood of \obj}

With a projected separation of 4100 AU from the four main components of
the \xiUMa\ system, \obj\ has a distant but interesting perspective.
The two binaries that form the 2\asec\ visual pair as seen from the
Earth are separated by 20 AU, or by an angular scale of 15\amin\
from the distant perspective of \obj.  At that distance the A and B components
would each shine with an apparent visual
magnitude of -9, one hundred times brighter than Venus in the skies of
Earth.

Given the complexity of the \xiUMa\ system, it is reasonable to
speculate that \obj\ could be a component that was once more
closely bound to the system, which was thrown into an orbit
with a very large apocenter during a three body interaction
that tightened the \xiUMa\  Bb binary.  
This would make \obj\ similar in history to an Oort Cloud comet,
formed at a radius of tens of AU from the central star system, lifted into
an orbit with a high apocenter by multi-body perturbations, followed
by galactic tides raising the pericenter \citep{duncan/quinn/tremaine:1987}.
Given an orbital period of order $10^5$ years,
the orbital eccentricity of \obj\ may ultimately be detectable.
With an apparent orbital radius of 8.5\amin\ the
orbital motion will amount to tens of milliarcseconds per year
relative to the primary system, and a radial velocity difference
of about 1 km/sec.  
Characterizing this motion,
specifically the differential proper motion between \obj\ and its
primary, is tractable given modern infrared and visual astrometric
capability, and certainly will be accomplished with the passage of time.

\section{Conclusions}

Alula Australis (\xiUMa), a solar neighborhood visual binary where
each component is itself a spectroscopic binary, possesses an ultra-cool
brown dwarf (T8.5) companion at a projected separation of 4100 AU.
This system is similar to, but somewhat more complex than, the
binaries studied by \citet{allen/etal:2012}, who found that 1 in 5
spectroscopic binary systems had distant common proper motion companions.
\citet{faherty/etal:2010}
found that wide companions were much more likely with binary or tertiary
central objects, and \xiUMa\ is a continuation of that trend.
Thus the \xiUMa\ system is uncommon primarily in the apparent magnitude
of the central star system, due to its proximity to the Sun.
The \xiUMa\  system provides a very accurate absolute magnitude
for a T8.5 brown dwarf, but both the age of the system and the
mass of the brown dwarf are uncertain due to the difficulty of
determining the age of a star in the middle of its main sequence
lifetime.
\obj\ appears to be slightly less luminous and redder than the
T8 dwarf WISEP J1423+01 which is a common proper motion companion
to BD+01 2920 \citep{pinfield/etal:2012}.

The \xiUMa\ system is amenable to detailed study since it is quite close
to the Sun.
An accurate (sub km/sec) radial velocity for \obj\ would be valuable information for
determining its orbit around \xiUMa.  
Improved proper motions for
both \obj\ and the center of mass of the central quadruple star system are
also needed for orbit characterization.

There is a tension between the red J-W2 color of \obj\ and its absolute magnitude
$M_{W2}$ when fitting to older models.
This discrepancy could be relaxed if \obj\  were a binary, but one AO observation
showed no evidence for binarity.
Thus the new types of clouds considered by \citet{morley/etal:2012} may well be
significant, and we conclude
that model spectra need to be updated to fit the observations of \obj\ and the many
other late T and Y dwarfs found in the WISE data.  

\acknowledgments

This publication makes use of data products from the Wide-field Infrared Survey Explorer, 
which is a joint project of the University of California, Los Angeles, and the 
Jet Propulsion Laboratory/California Institute of Technology, funded by the 
National Aeronautics and Space Administration.

This work was partially supported by a NASA Keck PI Data Award, administered by the NASA 
Exoplanet Science Institute. Part of the data presented herein were obtained at the W. M. Keck 
Observatory from telescope time allocated to the National Aeronautics and Space 
Administration through the agency's scientific partnership with the California Institute of 
Technology and the University of California. The Observatory was made possible by the 
generous financial support of the W. M. Keck Foundation.  The authors wish to recognize 
and acknowledge the very significant cultural role and reverence that the summit of Mauna Kea 
has always had within the indigenous Hawaiian community. We are most fortunate to have the 
opportunity to conduct observations from this mountain.

{\it Facilities:} \facility{WISE}, \facility{Fan Mountain/FanCam}, \facility{LBT/LUCIFER}, \facility{Palomar/TripleSpec},\\
\facility{Mt. Bigelow/Kuiper-2MASS}, \facility{WIYN/WIRC}, \facility{Keck/NIRC2-LGSAO}.

\end{document}